\documentclass{article}
\usepackage{spconf,amsmath,graphicx}

\usepackage{mathtools}
\usepackage{graphics, theorem, amsfonts, graphicx, amssymb, cite}
\usepackage{color}
\usepackage{setspace}
\usepackage{multirow}
\usepackage{rotating}
\usepackage{comment}
\usepackage[font=footnotesize]{caption}
\usepackage{algorithm}
\usepackage{algorithmic}
\usepackage{enumerate}
\usepackage{cite}
\usepackage{booktabs}
\usepackage{subcaption}
\PassOptionsToPackage{hyphens}{url}
\usepackage{hyperref}

\title{Wireless Link Scheduling via Graph Representation Learning: \\A Comparative Study of Different Supervision Levels}
\name{Navid Naderializadeh}
\address{University of Pennsylvania}

\begin{document}
\setlength{\abovedisplayskip}{5pt}
\setlength{\belowdisplayskip}{5pt}
%
\maketitle
\begin{abstract}
We consider the problem of binary power control, or link scheduling, in wireless interference networks, where the power control policy is trained using graph representation learning. We leverage the interference graph of the wireless network as an underlying topology for a graph neural network (GNN) backbone, which converts the channel matrix to a set of node embeddings for all transmitter-receiver pairs. We show how the node embeddings can be trained in several ways, including via supervised, unsupervised, and self-supervised learning, and we compare the impact of different supervision levels on the performance of these methods in terms of the system-level throughput, convergence behavior, sample efficiency, and generalization capability.\footnote{Code available at \url{https://github.com/navid-naderi/LinkSchedulingGNNs_SupervisionStudy}.}
\end{abstract}
\section{Introduction}
\label{sec:intro}

Over the past few years, machine learning, and particularly deep learning, architectures have been increasingly used to address challenging problems in wireless communications, including those that fall under the umbrella of radio resource management (RRM), such as beamforming and power control~\cite{sun2017learning, shen2021ai, shen2020graph}. More recently, solutions based on graph neural networks (GNNs), or in general, graph representation learning, have become more popular, mainly due to their desirable properties, such as permutation equivariance, size invariance, and stability to perturbations~\cite{shen2019graph, eisen2020optimal, he2021overview, gama2020stability}.

\begin{figure}[t]
\centering
\includegraphics[width=.35\textwidth]{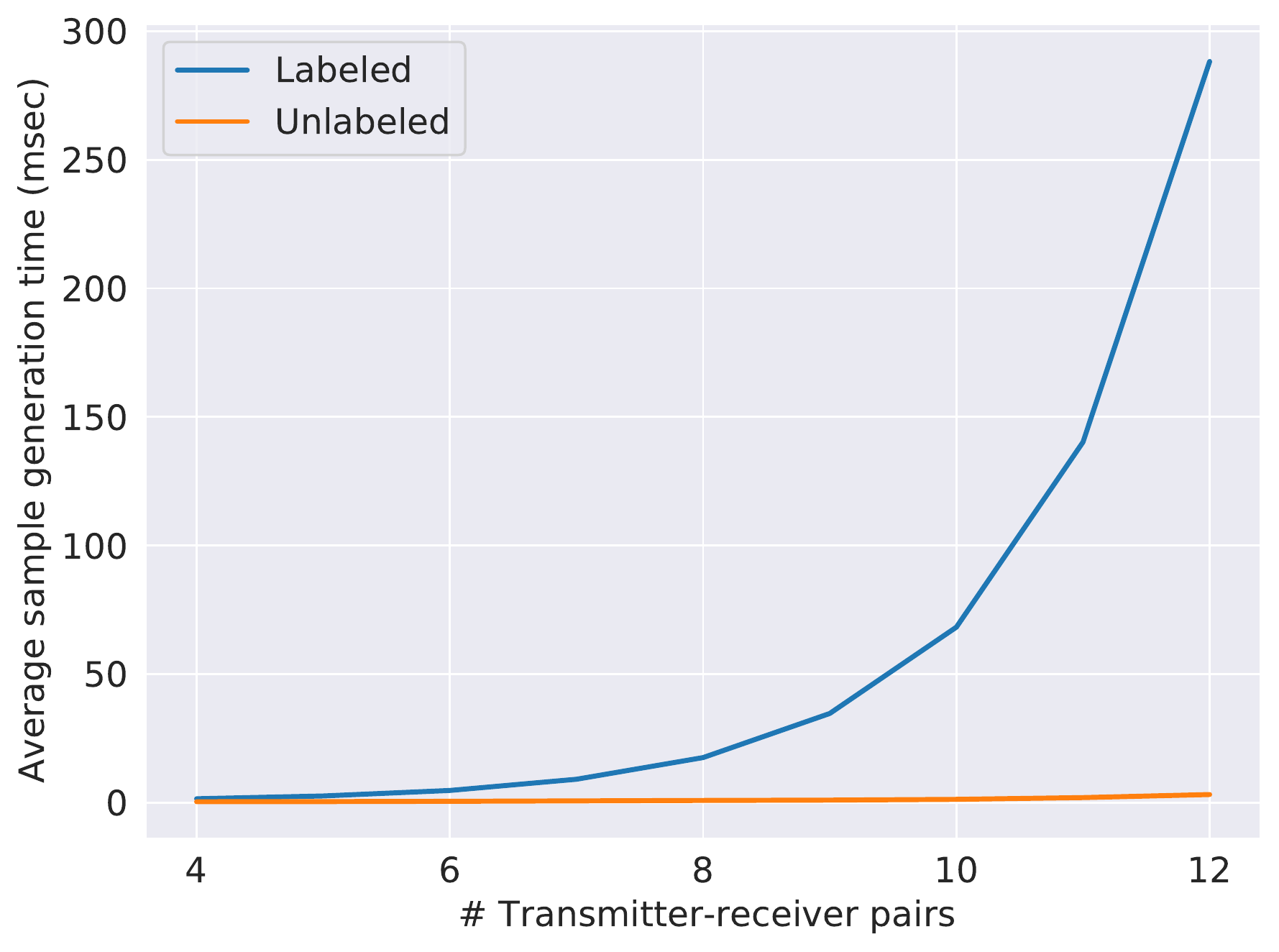}
\caption{Average wall-clock time to generate a single unlabeled/labeled sample for link scheduling in wireless networks with multiple transmitter-receiver pairs, where the labels are generated using exhaustive search.}
\label{fig:sample_generation_time}
\end{figure}

Nevertheless, it has been shown in~\cite{song2021supervise} that when using supervised learning for RRM problems, the \emph{quality} of the labels has a significant impact on the performance and convergence of the resulting models. For a special case of RRM problems which we focus on in this paper, i.e., link scheduling or binary power control, the highest-quality labels can be derived using exhaustive search. Figure~\ref{fig:sample_generation_time} compares the average time it takes to generate an unlabeled sample, i.e., the channel matrix, and a labeled sample, including the optimal link scheduling decisions generated by exhaustive search. It is clear that as the network size grows, generating high-quality labeled samples becomes exponentially more costly. This computational complexity has led to alternatives to supervised learning for training deep learning models in RRM problems, including unsupervised learning,  reinforcement learning, self-supervised learning, and meta-learning, which do not necessarily rely on (extensive) labeling of the data for training the underlying neural networks~\cite{shen2019graph, eisen2019learning, eisen2020optimal, naderializadeh2020wireless, lee2020graph, shen2020graph, nasir2019multi, naderializadeh2021resource, RRM_SSL_ICASSP2021, nikoloska2021fast, nikoloska2021black}. However, except for a few recent studies, such as~\cite{lee2020graph, shen2020graph, song2021supervise}, little effort has been made to thoroughly compare the performance of models trained using supervised and unsupervised learning procedures.

In this paper, we consider graph representation learning algorithms for the problem of link scheduling in wireless networks with multiple transmitter-receiver pairs, and study the impact of supervision level and type on the model performance using various metrics. We specifically use a GNN module to operate on the network graph in order to derive a set of \emph{node embeddings} for all transmitter-receiver pairs in the network, which are then used to learn optimal link scheduling decisions~\cite{lee2020graph}. We show how such decisions can be learned in both supervised and unsupervised learning, and we compare the resulting models in terms of system-level throughput, convergence behavior, sample complexity, and generalization capability. We also show how self-supervised learning can be used to pre-train the GNN backbones, further boosting the resultant supervised/unsupervised models~\cite{RRM_SSL_ICASSP2021}.

\section{Problem Formulation}\label{sec:problem_formulation}
We consider a $K$-user interference channel with $K$ transmitter-receiver pairs $\{(\mathsf{Tx}_i, \mathsf{Rx}_i)\}_{i=1}^K$, where each transmitter $\mathsf{Tx}_i$ intends to communicate with its corresponding receiver $\mathsf{Rx}_i$. We assume that simultaneous transmissions of multiple transmitters cause interference on each other, as they all use the same time/frequency/spatial resources. Let $h_{ij}\in\mathbb{C}$, $P_{\max}$, and $N$ denote the channel gain between transmitter $\mathsf{Tx}_j$ and receiver $\mathsf{Rx}_i$, maximum possible transmit power, and Gaussian noise variance at each receiver, respectively. Then, assuming that each receiver treats the interference from other transmitters as noise~\cite{geng2015optimality}, 
the Shannon capacity of the channel between transmitter $\mathsf{Tx}_i$ and receiver $\mathsf{Rx}_i$ is given by
\begin{align}
R_i &= \log_2\left(1 + \tfrac{|h_{ii}|^2 \gamma_i}{\sum_{{j=1, j\neq i}}^K |h_{ij}|^2 \gamma_j + \frac{N}{P_{\max}}}\right),
\end{align}
where $\gamma_i=\frac{P_i}{P_{\max}}\in[0,1]$ denotes the normalized transmit power used by transmitter $\mathsf{Tx}_i$.

In this paper, we focus on binary power control, i.e., \emph{link scheduling}, where $\gamma_i\in \{0,1\}, \forall i\in\{1,\dots,K\}$, for maximizing the sum-rate across the network. In particular, we intend to solve the following non-convex optimization problem:
\vspace{-.13in}
\begin{subequations}\label{eq:opt_PC}
\begin{alignat}{2}
&\max_{\gamma_1,...,\gamma_K}&& \quad \sum_{i=1}^K \log_2\left(1 + \tfrac{|h_{ii}|^2 \gamma_i}{\sum_{{j=1, j\neq i}}^K |h_{ij}|^2 \gamma_j + \frac{N}{P_{\max}}}\right)\label{eq:obj_function} \\
&\quad~\text{s.t.} && \quad \gamma_i\in\{0, 1\}, ~\forall i\in\{1,\dots,K\}.
\end{alignat}
\end{subequations}

\section{Link Scheduling via Graph~Representation~Learning}
\vspace{-.1in}
To apply graph representation learning approach to the link scheduling problem in~\eqref{eq:opt_PC}, similar to~\cite{lee2020graph, shen2020graph}, we first represent a given $K$-user interference channel with channel matrix $\mathbf{H}\in\mathbb{C}^{K\times K}$ as a directed graph $G^{\mathbf{H}}=(\mathcal{V}^{\mathbf{H}}, \mathcal{E}^{\mathbf{H}}, \alpha^{\mathbf{H}}, \beta^{\mathbf{H}})$, where $\mathcal{V}^{\mathbf{H}}=\{1,\dots,K\}$ denotes the set of graph nodes, where the $i$\textsuperscript{th} node corresponds to the $i$\textsuperscript{th} transmitter-receiver pair $(\mathsf{Tx}_i, \mathsf{Rx}_i)$. The set of edges in the graph is denoted by $\mathcal{E}^{\mathbf{H}}=\{(u, v)\in\mathcal{V}^{\mathbf{H}}\times\mathcal{V}^{\mathbf{H}}: u \neq v \}$. Furthermore, $\alpha^{\mathbf{H}}:\mathcal{V}^{\mathbf{H}} \rightarrow \mathbb{R}^{F_0}$ and $\beta^{\mathbf{H}}:\mathcal{E}^{\mathbf{H}} \rightarrow \mathbb{R}$ denote functions which determine initial node feature vectors $\mathbf{x}^0_v=\alpha^{\mathbf{H}}(v), \forall v\in\mathcal{V}$ and edge weights $e_{u, v}=\beta^{\mathbf{H}}(u, v), \forall (u, v)\in\mathcal{E}$, respectively. In the following, we drop the dependence of the graph $G$ on $\mathbf{H}$ for brevity unless necessary.

The aforementioned graph $G$ serves as the underlying topology for a graph neural network (GNN) backbone, which takes as input the graph and produces as output a \emph{node embedding} for each node in the graph. In particular, the GNN processes the set of node features through a sequence of $L$ layers, where at each layer $l\in\{1,\dots,L\}$, the feature vector of each node $v\in\mathcal{V}$ is updated as
\begin{align}\label{eq:gnn_combine}
\mathbf{x}_{v}^l = \phi_l\left(\mathbf{x}_{v}^{l-1}, \left\{\mathbf{x}_{u}^{l-1}, e_{u, v}\right\}_{u\in\mathcal{V}: (u, v)\in \mathcal{E}}\right),
\end{align}
where $\phi_l$ denotes a parametric combining function, whose parameters are learned, e.g., through backpropagation of the gradients of an objective function. This implies that each node combines its own features from the previous layer, the features of its incoming neighbors from the previous layer, and the incoming edge weights into a new feature vector $\mathbf{x}_{v}^l \in \mathbb{R}^{F_l}$. The resulting feature vectors of each node at the end of $L$ layers, i.e., $\left\{\mathbf{x}_{v}^L\right\}_{v\in\mathcal{V}}$, are called \emph{node embeddings}.

Once the node embeddings are created, they finally undergo a \emph{link scheduling head}, denoted by $\psi:\mathbb{R}^{F_L} \rightarrow [0, 1]$, which is another parametric function that maps each node embedding $\mathbf{x}_{v}^L$ to a normalized power level $\psi(\mathbf{x}_{v}^L)$ for the corresponding transmitter $\mathsf{Tx}_v$. These continuous normalized power levels can be translated into link scheduling decisions using a thresholding mechanism, e.g., $\gamma_v=\mathbb{I}(\psi(\mathbf{x}_{v}^L) \geq 0.5)$, where $\mathbb{I}(\cdot)$ denotes the indicator function.

\subsection{Supervised and Unsupervised Training}
The parameters of the GNN backbone and the link scheduling head can be trained in an end-to-end manner using either supervised or unsupervised learning. Assume that we have access to a batch of $B$ labeled samples $\{(G_i, \mathbf{\Gamma}^*_i)\}_{i=1}^B$, where for a given network graph $G_i=(\mathcal{V}_i, \mathcal{E}_i, \alpha_i, \beta_i)$, $\mathbf{\Gamma}^*_i=(\gamma^*_{i,1}, \dots, \gamma^*_{i,|\mathcal{V}_i|})$ denotes the optimal link scheduling decisions derived using exhaustive search. Then, the supervised loss can be written as the cross-entropy loss between the model outputs and the optimal power control decisions,
\begin{align}\label{eq:supervised_loss_main}
\mathcal{L}_{\mathsf{supervised}} &= -\frac{1}{B} \sum_{i=1}^B \sum_{v\in\mathcal{V}_i} \ell_{i, v},
\vspace{-.15in}
\end{align}
where we define
\begin{align}\label{eq:supervised_loss_2}
\ell_{i, v}&= \gamma^*_{i,v} \log_2(\psi(\mathbf{x}_{v}^L)) + (1-\gamma^*_{i,v}) \log_2(1-\psi(\mathbf{x}_{v}^L)).
\end{align}

On the other hand, an unsupervised training procedure directly tunes the model parameters to maximize the objective function in~\eqref{eq:obj_function}, without the need for any ground-truth labels, i.e., optimal power control decisions. Specifically, given a batch of $B$ unlabeled samples $\{G_i^{\mathbf{H}_i}\}_{i=1}^B$, the unsupervised loss function is given by
\vspace{-.15in}
\begin{align}\label{eq:unsupervised_loss_main}
\mathcal{L}_{\mathsf{unsupervised}} &= -\frac{1}{B} \sum_{i=1}^B \sum_{v\in\mathcal{V}_i} r_{i, v},
\end{align}
where we have
\begin{align}\label{eq:unsupervised_loss_2}
r_{i, v} = \log_2\left(1 + \tfrac{|h_{i, vv}|^2 \psi(\mathbf{x}_{v}^L)}{\sum_{u\in\mathcal{V}_i\setminus\{v\}} |h_{i, vu}|^2 \psi(\mathbf{x}_{u}^L) + \frac{N}{P_{\max}}}\right).
\end{align}

\subsection{Self-Supervised Pre-Training}
It was shown in~\cite{RRM_SSL_ICASSP2021} that using contrastive self-supervised pre-training (as originally proposed in~\cite{oord2018representation, chen2020simple}) helps significantly improve the sample efficiency of a subsequent supervised training process for learning power control decisions. Assuming a batch of unlabeled \emph{augmented} graph pairs $\{(\overline{G}_i, \underline{G}_i)\}_{i=1}^B$, where for any $i\in\{1,\dots,B\}$, $\overline{G}_i$ and $\underline{G}_i$ are \emph{semantically similar}, the self-supervised loss is given by
\vspace{-.05in}
\begin{align}\label{eq:contrastive_loss_main}
\setlength{\belowdisplayskip}{0pt}
\setlength{\belowdisplayshortskip}{0pt}
\mathcal{L}_{\mathsf{self}\text{-}\mathsf{supervised}} &= -\frac{1}{B} \sum_{i=1}^B \sum_{v\in\mathcal{V}_i} \log_2 \left(\frac{e^{\frac{(\overline{\mathbf{x}}_{v}^L)^T \underline{\mathbf{x}}_{v}^L}{\tau}}}{\sum\limits_{j=1}^B \sum\limits_{u\in\mathcal{V}_j} e^{\frac{(\overline{\mathbf{x}}_{v}^L)^T \underline{\mathbf{x}}_{u}^L}{\tau}}} \right),
\vspace{-.05in}
\end{align}
where $\left\{\overline{\mathbf{x}}_{v}^L\right\}_{v\in\overline{\mathcal{V}}_i}$ and $\left\{\underline{\mathbf{x}}_{v}^L\right\}_{v\in\underline{\mathcal{V}}_i}$ respectively denote the set of node embeddings when the GNN backbone is applied to $\overline{G}_i$ and $\underline{G}_i$, and $\tau$ denotes a temperature hyperparameter. Note that as the self-supervised loss~\eqref{eq:contrastive_loss_main} is defined on the node embeddings, it only impacts the GNN backbone parameters. The parameters of the link scheduling head need to be trained using either the supervised loss~\eqref{eq:supervised_loss_main} or the unsupervised loss~\eqref{eq:unsupervised_loss_main}.

\begin{figure*}[t!]
    \centering
    \begin{subfigure}[t]{0.24\textwidth}
        \centering
        \includegraphics[width=\linewidth]{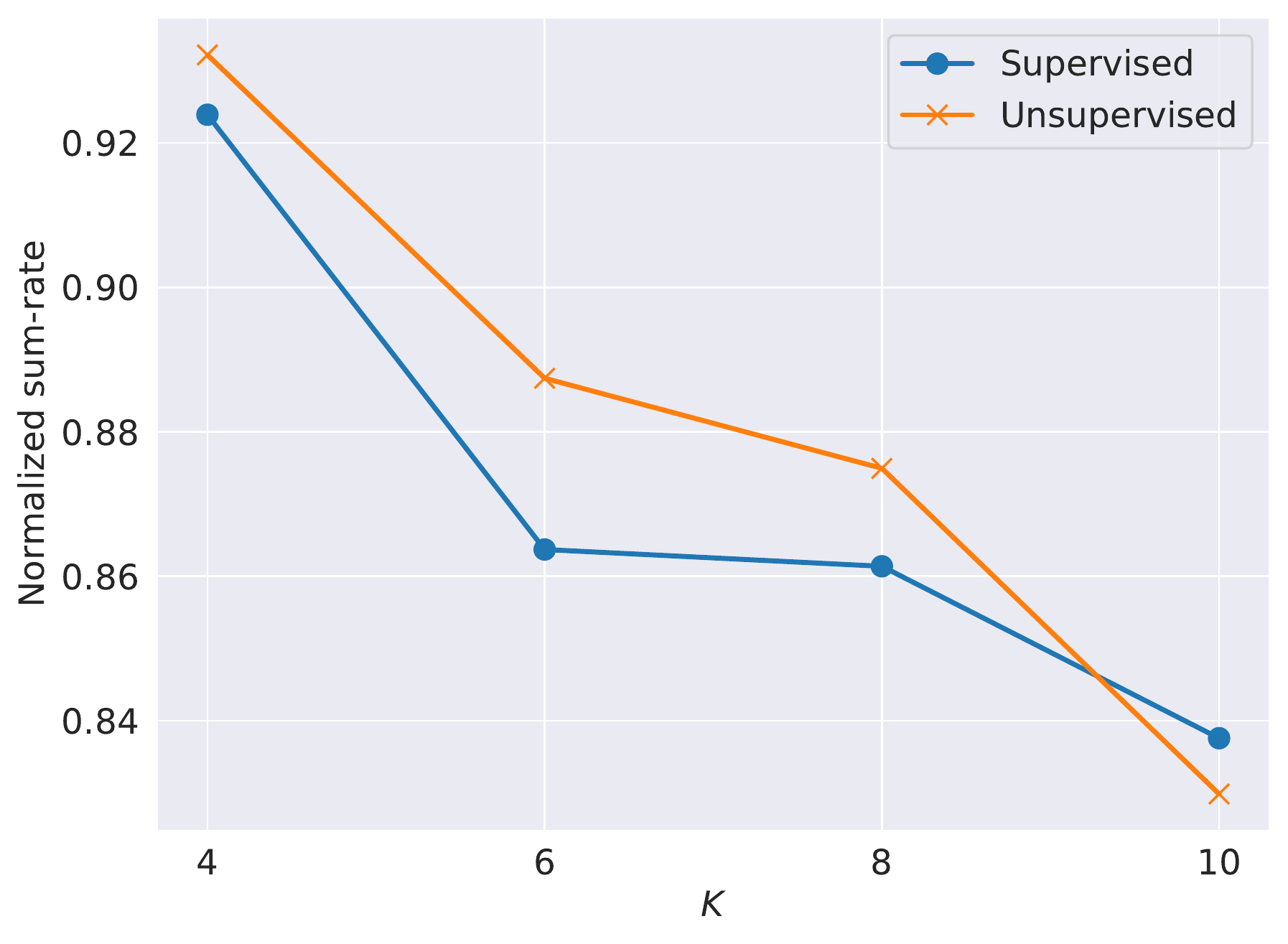}
        \caption{}
        \label{fig:throughput_vs_k}
    \end{subfigure}%
    ~ 
    \begin{subfigure}[t]{0.24\textwidth}
        \centering
        \includegraphics[width=\linewidth]{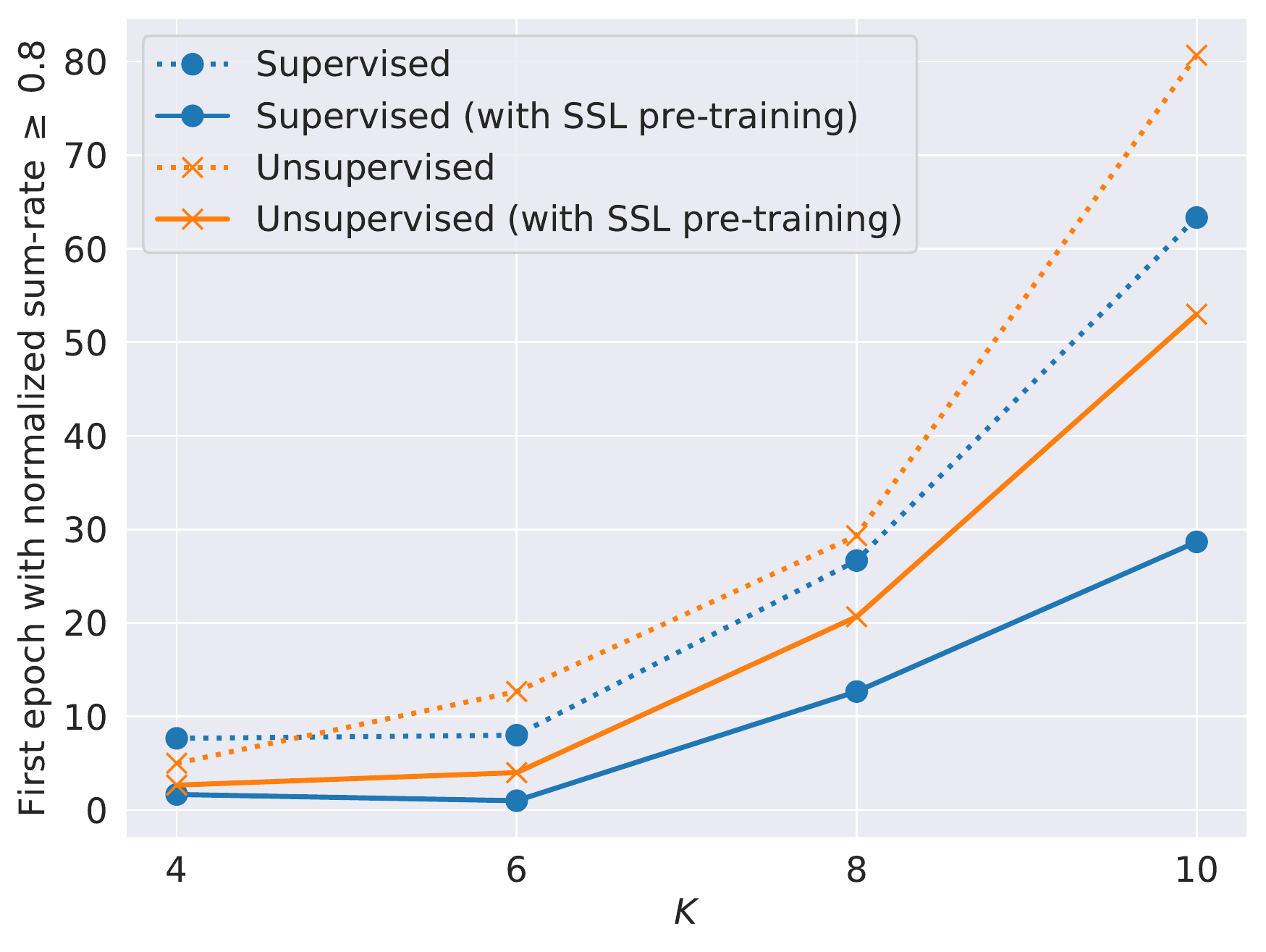}
        \caption{}
        \label{fig:convergence_vs_k}
    \end{subfigure}%
    ~ 
    \begin{subfigure}[t]{0.24\textwidth}
        \centering
        \includegraphics[width=\linewidth]{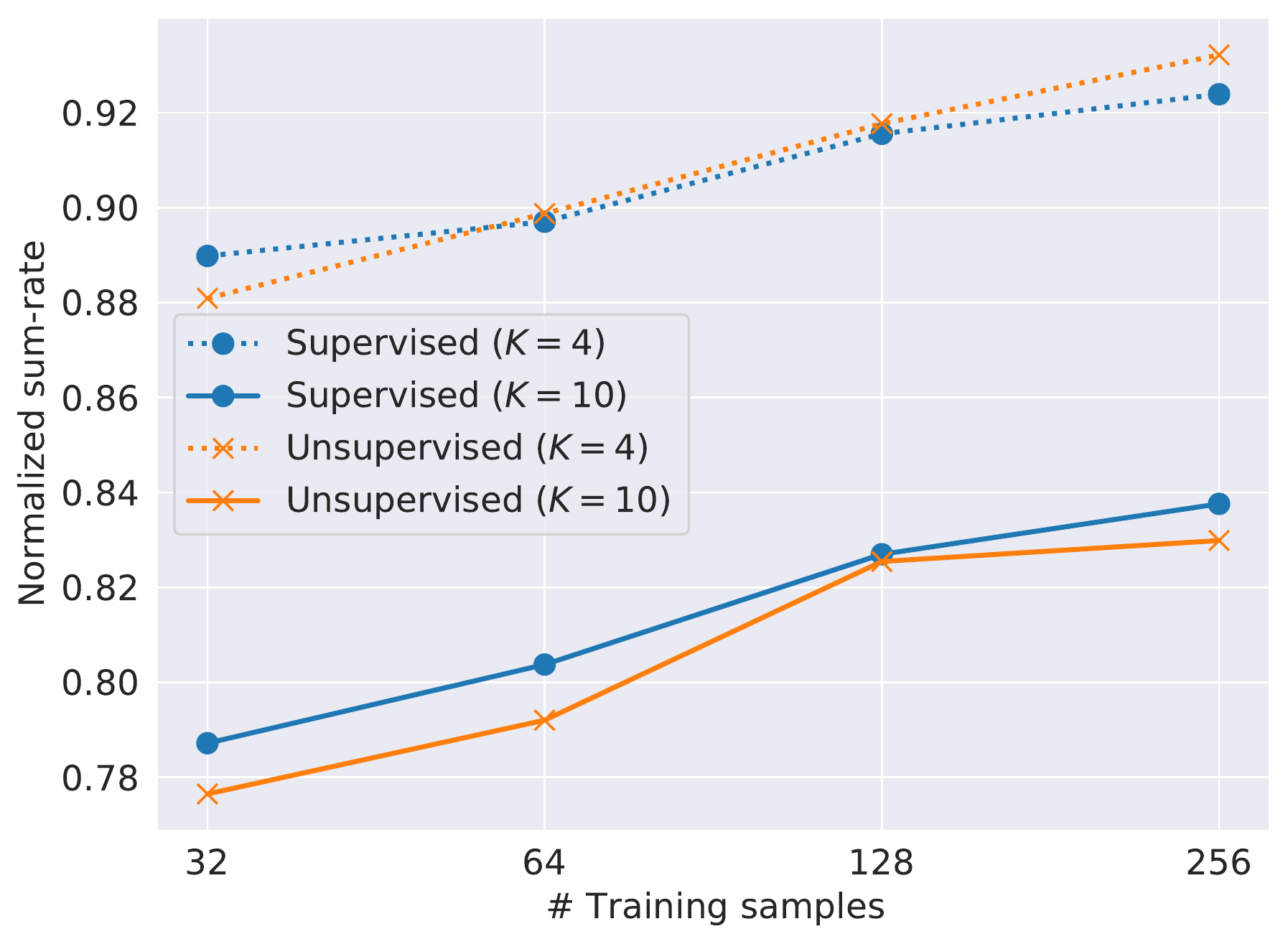}
        \caption{}
        \label{fig:sample_complexity}
    \end{subfigure}%
    ~ 
    \begin{subfigure}[t]{0.24\textwidth}
        \centering
        \includegraphics[width=\linewidth]{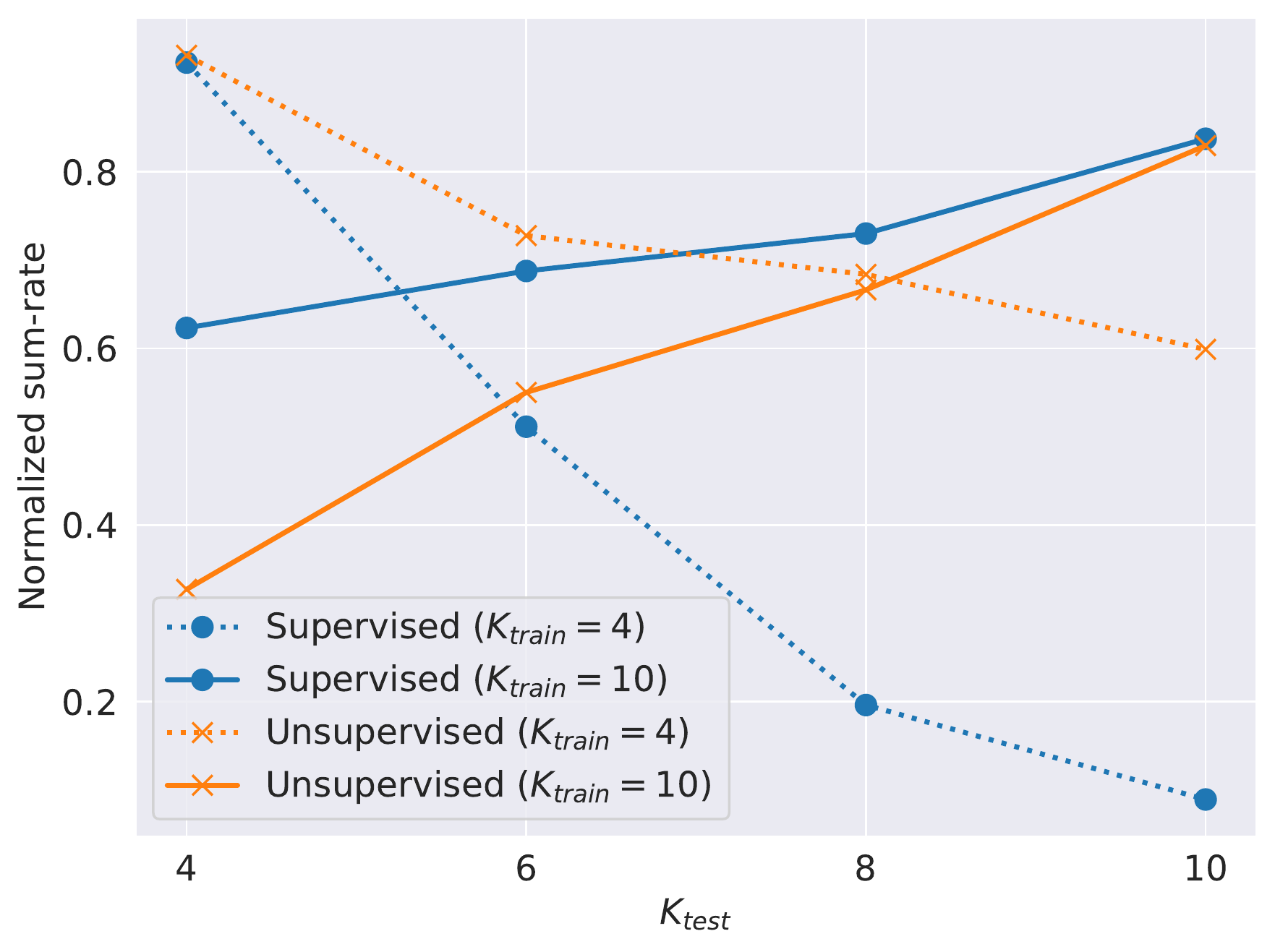}
        \caption{}
        \label{fig:generalization_vs_k}
    \end{subfigure}
    \caption{Comparison of supervised and unsupervised models for different network sizes in terms of (a) normalized sum-rate, (b) convergence behavior with/without self-supervised pre-training (SSL), (c) sample complexity, and (d) generalization capability.}
    \label{fig:kshot_sumrates}
\end{figure*}

\section{Empirical Evaluation of the Impact of Supervision Level on Trained Models}
Given that the cost of optimal labeling of samples in a link scheduling problem grows exponentially with the network size (see Figure~\ref{fig:sample_generation_time}), it is imperative to understand how the supervised and unsupervised losses in~\eqref{eq:supervised_loss_main} and~\eqref{eq:unsupervised_loss_main} compare to each other in different aspects. While some initial comparative studies have been conducted in prior work, e.g., in~\cite{lee2020graph, shen2020graph, song2021supervise}, in this section we perform a more comprehensive analysis to study how these different supervision levels impact the resulting models in terms of i) system sum-rate, ii) convergence behavior, iii) sample complexity, and iv) generalization capability. We also show how self-supervised pre-training can help improve models trained under both supervised and unsupervised loss types.

We consider $K$-user interference channels with $K\in\{4, 6, 8, 10\}$, and for each value of $K$, we generate a total of $256$ training samples and $256$ testing samples, all labeled using exhaustive search. For each sample, the $K$ transmitters are located uniformly at random within a $250m \times 250m$ area, ensuring a minimum inter-transmitter distance of $35m$. Then, for each transmitter, the corresponding receiver is located uniformly at random inside a ring around the transmitter, with inner and outer radii of $10m$ and $50m$, respectively. We follow the dual-slope path-loss model of~\cite{naderializadeh2020wireless, naderializadeh2021resource} and a log-normal shadowing with $7$dB standard deviation to generate the channel gains. We set the maximum transmit power, bandwidth, and noise power spectral density to $P_{\max}=10$dBm, $10$MHz, and $-174$dBm/Hz, respectively.

For the graph generation, we use scalar initial node features (i.e., $F_0=1$), where for a node $v\in\mathcal{V}$, we set
\begin{align}
\mathbf{x}^0_v=\log\left(\frac{P_{\max}|h_{vv}|^2}{N}\right) / Z,
\end{align}
where $Z$ is a normalization factor, defined as
\begin{align}
Z=\left(\sum_{(u,v)\in\mathcal{V} \times \mathcal{V}}\left|\log\left(\frac{P_{\max}|h_{vu}|^2}{N}\right)\right|^2\right)^{\frac{1}{2}}.
\end{align}
This implies that the initial feature of each node is the normalized signal-to-noise ratio (SNR) between the corresponding transmitter-receiver pair (in dB). Similarly, for each edge $(u,v)\in\mathcal{E}$, we define its edge weight as the normalized interference-to-noise ratio (INR) from transmitter $\mathsf{Tx}_u$ to receiver $\mathsf{Rx}_v$ (in dB), i.e.,
\begin{align}
e_{u,v}=\log\left(\frac{P_{\max}|h_{vu}|^2}{N}\right) / Z.
\end{align}
As for the GNN, we use the local extremum operator proposed in~\cite{ranjan2020asap}, where the combining function~\eqref{eq:gnn_combine} is given by
\vspace{.05in}
\begin{align}
\mathbf{x}_{v}^l = \mu\left(\mathbf{x}_{v}^{l-1} \mathbf{\Theta}_{l,1} + \hspace{-.08in} \sum_{u: (u, v)\in \mathcal{E}} \hspace{-.07in} e_{u, v} \left(\mathbf{x}_{v}^{l-1} \mathbf{\Theta}_{l,2} - \mathbf{x}_{u}^{l-1} \mathbf{\Theta}_{l,3} \right)\right).
\end{align}
Here, $\mathbf{\Theta}_{l,1}, \mathbf{\Theta}_{l,2}, \mathbf{\Theta}_{l,3}$ denote learnable parameters, all residing in $\mathbb{R}^{F_{l-1} \times F_l}$, and $\mu(\cdot)$ represents a LeakyReLU non-linearity (with a negative slope of $10^{-2}$). We use $L=3$ hidden layers and set $F_1=F_2=F_3=64$. Moreover, we parameterize the link scheduling head as
\begin{align}
\psi(\mathbf{x}_{v}^L) = \sigma\left(\mathbf{w}^T \mathbf{x}_{v}^L + b \right),
\end{align}
where $\mathbf{w}\in\mathbb{R}^{F_L}$ and $b\in\mathbb{R}$ represent learnable parameters, and $\sigma(\cdot)$ denotes the sigmoid function.

We use a learning rate of $10^{-2}$, set the batch size to $B=32$, train each model for $500$ epochs, and report the maximum test sum-rate normalized by the optimal sum-rate under binary power control (derived using exhaustive search). For models with self-supervised pre-training, we train them for $100$ epochs using the contrastive loss in~\eqref{eq:contrastive_loss_main} with $\tau=0.1$ before the main supervised/unsupervised training phase. To create augmentations, we use a similar process as in~\cite{RRM_SSL_ICASSP2021} based on the information-theoretic optimality condition of treating interference as noise~\cite{geng2015optimality, naderializadeh2014itlinq}, coupled with multiplicative perturbations of the channel gains, drawn uniformly at random from the interval $[0.9, 1.1]$. We run each experiment with three different random seeds and report the mean performance across the three resulting models. All the training and testing procedures are implemented using the PyTorch Geometric library~\cite{Fey/Lenssen/2019}. Unless explicitly stated, we assume each model is trained and tested on a fixed network size, i.e., $K$.

\subsection{System Sum-Rate}


Figure~\ref{fig:throughput_vs_k} shows the normalized test sum-rates achieved by the supervised and unsupervised models for different numbers of transmitter-receiver pairs. As the figure demonstrates, both types of models achieve more than $80\%$ of the optimal sum-rate of exhaustive search. However, as expected, the normalized sum-rates of both model types decline with increased network size, which leads to more complex link scheduling decisions since the density of the links grows in the fixed network area. Interestingly, unsupervised models outperform the supervised models in smaller networks, but they are overtaken by the supervised models for $K=10$, potentially since supervision is more helpful in larger network sizes due to more complex interference patterns.

\subsection{Convergence Behavior}


In Figure~\ref{fig:convergence_vs_k}, we compare the supervised and unsupervised models in terms of their convergence behavior. In particular, we plot the first epoch in which each model's normalized test sum-rate exceeds $0.8$. As expected, both models take longer to converge with increased network size. However, without any supervision, unsupervised models take significantly longer to converge as compared to supervised models, especially for larger network sizes. As the plot shows, self-supervised pre-training helps both supervised and unsupervised models converge considerably faster, with the supervised models exceeding $0.8$ normalized sum-rate within the first \emph{two} epochs for $K\in\{4, 6\}$.

\subsection{Sample Complexity}


Another important aspect to study for the trained models is their sample complexity. Figure~\ref{fig:sample_complexity} illustrates the performance of supervised and unsupervised models for $K\in\{4, 10\}$, when the number of training samples changes from $32$ samples to the entire training set of $256$ samples. As the figure shows, for $K=4$, unsupervised models outperform supervised models with as few as 64 training samples. However, for $K=10$, supervised models outperform unsupervised models in the entire range of the training set size.

\subsection{Generalization Capability}


One of the main benefits of GNNs is their size-invariance, implying that a model trained on a specific network size, $K_{\emph{train}}$, can be evaluated on any other network size $K_{\emph{test}}$. Figure~\ref{fig:generalization_vs_k} shows how models trained on $K_{\emph{train}}\in\{4, 10\}$ perform on networks with $K_{\emph{test}}\in\{4, 6, 8, 10\}$. As the figure shows, the performance of both supervised and unsupervised models is highest when $K_{\emph{test}}=K_{\emph{train}}$. However, depending on the value of $K_{\emph{train}}$, supervised and unsupervised models show significantly different behaviors: Unsupervised models trained on small networks with $K_{\emph{train}}=4$ demonstrate considerably better generalization capability as compared to supervised models. However, when trained on large networks with $K_{\emph{train}}=10$, supervised models generalize much better than unsupervised models. This again shows that as the network gets denser and more complex interference patterns emerge, supervised training leads to better models, but at the expense of the costly labeling process.

\section{Discussion and Concluding Remarks}
In this paper, we studied learning-based solutions for link scheduling in wireless networks using graph representation learning. We considered three types of loss functions with different supervision levels, including supervised, unsupervised, and self-supervised losses, and compared the trained graph-neural-network-based models using these losses from different angles. We showed that unsupervised models generally outperform supervised models when trained on smaller network sizes. However, supervised models are superior in terms of system-level performance, convergence behavior, sample efficiency, and generalization capability when trained on denser networks, though this comes at the exponentially high cost of obtaining high-quality labeled samples in these large configurations. We further showed how self-supervised pre-training can boost both supervised and unsupervised models in terms of convergence behavior. Our results confirm the findings of prior work in terms of the trade-offs between the considered metrics for models trained using supervised and unsupervised learning. The very feasibility of unsupervised learning approaches hinges on having the complete knowledge of the objective function. Therefore, in radio resource management problems where the objective function is unknown or not differentiable, unsupervised learning is not possible, and approaches based on reinforcement learning can instead be applied. Comparing the performance of reinforcement learning and supervised learning approaches in such scenarios, especially for modeling the temporal aspects of the underlying problems, is an interesting research topic, which we leave for future work.


\bibliographystyle{IEEEbib}
\bibliography{references}

\end{document}